\documentclass[%
 reprint,
 amsmath,amssymb,
 aps,
]{revtex4-2}

\usepackage{graphicx}
\usepackage{dcolumn}
\usepackage{bm}
\usepackage{subfigure}
\usepackage{float}

\begin{document}

\preprint{APS/123-QED}

\title{Design, fabrication and test of parallel-coupled slow-wave high-gradient structure for ultrashort input power pulses}

\author{Weihang Gu, Hao Zha\thanks{zha@tsinghua.edu.cn}, Jiaru Shi ,Yuliang Jiang, Jiayang Liu, Xiancai Lin, Focheng Liu, and Huaibi Chen}

\affiliation{%
 Department of Engineering Physics, Tsinghua University, Beijing CN-100084, China. Key Laboratory of Particle and Radiation Imaging, Tsinghua University, Ministry of Education, Beijing CN-100084, China
}%

\date{\today}

\begin{abstract}
Tsinghua University has designed an X-band (11.424 GHz) slow-wave parallel-coupling accelerating structure, and demonstrated the high performance of the over-coupled structure operating with ultrashort pulse. In this study, we redesigned a 40-ns structure with 10 cells, tailored to the specifications of a high-power experimental platform, and provided a detailed analysis of the experimental results. Unexpected bead-pull results were observed during the cold testing, which we attribute to inter-cavity coupling. To explain these results, a multi-cell coupling circuit model was proposed and analyzed. High-power testing was conducted on the TPOT-X platform, and the highest gradient achieved was 130 MV/m after \textit{$1.1 \times 10^7$} conditioning pulses. Compared to conventional multi-cavity high-gradient structures, the distributed power feeding system offers a shorter conditioning period and demonstrates the potential to achieve higher accelerating gradients under short-pulse operation.
\end{abstract}

\maketitle


\section{introduction}

Conventional linear electric accelerators are extensively used in medical and industrial applications \cite{hanna2012rf,maxim2019phaser}, underscoring the importance of compact high-gradient accelerating structures. As a leading approach for generating high-energy electrons, high-gradient normal-conducting structures have been actively developed for use in colliders and light source facilities \cite{tang2009tsinghua,diomede2018preliminary,stragier2018smart}.For instance, CERN has designed a multi-cavity structure achieving a gradient of 120 MV/m for the Compact Linear Collider (CLIC)\cite{argyropoulos2018design}. Additionally, the main accelerating sections of free-electron lasers (FELs)\cite{PhysRevSTAB.17.080702} and VIGAS \cite{lin2022fabrication} employ multiple linacs operating across different frequency bands.

Despite ongoing advancements, the achievable accelerating gradient is fundamentally constrained by RF breakdown during the conditioning process\cite{shao2015observation,fowler1928electron,solyak2009gradient,wang2004accelerator,adolphsen2000rf,wang1997field}. Empirical data from CLIC experiments indicate that, at a fixed accelerating gradient, the breakdown rate (BDR) is proportional to the fifth power of the input pulse width\cite{grudiev2009new}. Therefore, shorter pulse is conducive to achieving higher accelerating gradients. Previous studies in the short-pulse regime have primarily focused on single cavities or high-frequency (30 GHz) experiments. Recently, an X-band traveling wave (TW) cavity structure tested at Argonne National Laboratory (ANL) demonstrated operation at gradients up to 313 MV/m with a pulse full-width at half-maximum (FWHM) of 6 ns \cite{shao2022demonstration}. For a 30 GHz TW structure, accelerating gradients of 140, 100, and 70 MV/m were reached with pulse widths of 4, 8, and 16 ns, respectively, at CLIC\cite{braun2001status,braun2003frequency}.

In conventional high-gradient linear accelerating structures, power is typically coupled axially through the beam aperture or coupling holes, constraining input pulse widths to the structure’s filling time\cite{higo2013comparison}.To obtain higher gradients, X-band traveling wave structures operate with a pulse width of more than 100 ns\cite{higo2010advances,degiovanni2016comparison}. Consequently, multi-cell structures designed to operate under short pulses are expected to further enhance accelerating gradients.

One approach to reduce the filling time is to fill cells in parallel, allowing power flow to bypass the resistance encountered in the cells. Based on this principle, the parallel-coupled accelerating structure was proposed as a topological configuration for linear accelerators. In this design, power is coupled into each cell independently through an external power distribution network \cite{sundelin1977parallel,ivannikov1986accelerating,neilson2011design}. Since RF power propagates through waveguides or coaxial lines at the speed of light, the electromagnetic fields in all cells are established simultaneously. This results in the overall filling time of the structure being nearly equal to that of a single cell. Additionally, this parallel power feeding method reduces the susceptibility of the structure to breakdown, as little to no power flows through the beam tunnel. It also provides greater flexibility in the geometric optimization of the cavity design \cite{tantawi2016distributed}. In 2018, SLAC developed the first high-gradient parallel-coupled 20-cell structure \cite{tantawi2018distributed}. This structure operates at 11.424 GHz and achieved an accelerating gradient of 140 MV/m under an input power of 33 MW with a 400 ns pulse width. However, this structure has not yet been further tested under short-pulse conditions.

\begin{figure}[!h]
\centering
\includegraphics[width=\columnwidth]{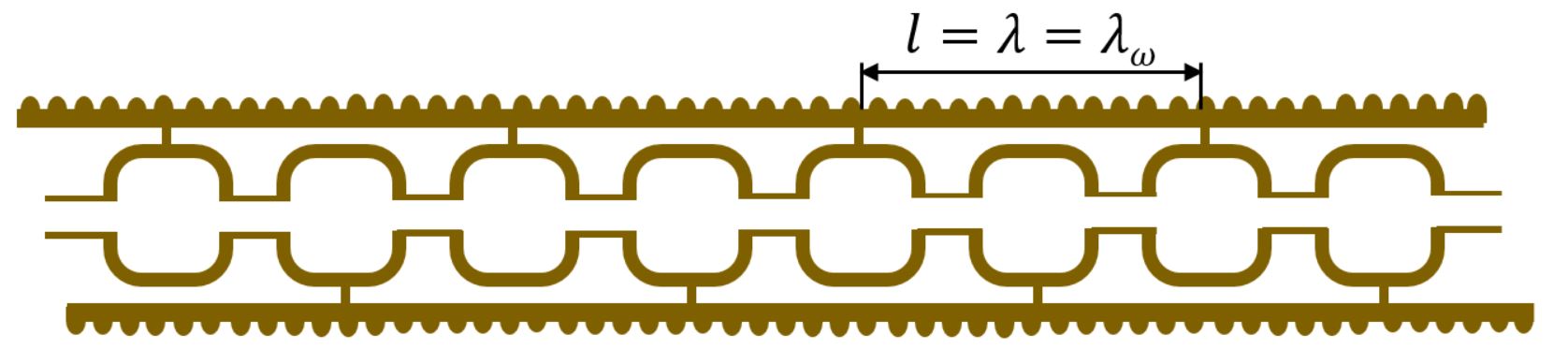}
\caption{ The slow-wave parallel-coupled accelerating structure designed by Tsinghua university.}
\label{slowwave_str}
\end{figure}

Tsinghua University has conducted research on the circuit design of parallel-coupled system networks and developed a short-pulse parallel-coupled accelerator structure \cite{jiang}, as shown in Fig.~\ref{slowwave_str}. In vacuum waveguides, the RF phase velocity exceeds the speed of light. To address this, periodic slots are employed as a slow-wave structure, ensuring proper phase matching for each cavity. Additionally, the phase advance is set to 0 or 180 degrees, ensuring equal power distribution across the cells. The entire structure is intentionally over-coupled to achieve a short filling time, with an effective coupling coefficient of approximately \textit{$\beta' =26$}. Theoretically, with a 10 ns pulse input at 1.1 GW, an average accelerating gradient of 200 MV/m can theoretically be achieved. This predicted result is derived from CLIC’s experimental findings and established empirical formulas.

To obtain valuable experimental data, this structure is planned for high-power testing on Tsinghua University’s X-band high-power testing platform, TPOT-X\cite{peng2018development}. TPOT-X operates at 11.424 GHz and employs a klystron as its power source. The platform can output up to 50 MW with a 1.5 $\mu$s microwave pulse at a maximum repetition rate of 40 Hz. With the assistance of a pulse compressor, TPOT-X is capable of generating microwave pulses with peak power in the hundreds of MW and pulse widths in the hundred-nanosecond range. However, the platform’s shortest achievable pulse width is approximately 40 ns due to limitations imposed by the pulse rising time and the need to maintain the performance of the pulse compressor. Therefore, the original 10-ns structure needs to be modified to satisfy the 40-ns pulse requirement.

In this paper, we redesigned, fabricated, and tested an X-band short-pulse parallel-coupled accelerating structure with 10 cells, based on previous research and within the limits of experimental conditions. In Sec.~\ref{rfdesign}, we present the RF design of the entire parallel-coupled accelerating structure and calculate the real-time field distribution of the structure. Sec.~\ref{tuning} describes the fabrication process and analyzes the results of the low-power tests. Sec.~\ref{test} reports the high-power experimental results of the structure under short-pulse conditions. Finally, Sec.~\ref{conclusion} provides a summary of the entire paper.

\section{\label{rfdesign}physics design and analysis}

Three subsections are presented in this section to describe the structure in detail. The first part focuses on the RF design of the slow-wave structure. In the second part, we describe the geometry of the accelerating cell. This includes an analysis of the single-cell design as well as the coupling coefficients of the entire structure. Finally, in the third part, we calculate the real-time distribution of the electromagnetic field in the parallel-coupled structure.

\subsection{distributing waveguide design}
The phase shift between two adjacent feeding ports is set to 0 or \textit{$\pi$} to ensure that each identical cell receives an equal amount of RF power. For electrons traveling at the speed of light within the structure, the RF phase velocity must be matched to the speed of light to achieve phase synchronization and maintain continuous acceleration. A smooth waveguide structure is unsuitable for this purpose because the guided wavelength, \textit{$\lambda _{g}$}, exceeds the free-space wavelength, \textit{$\lambda $}. To address this, periodic slots were introduced as a slow-wave structure, as shown in Fig.~\ref{slowwave_str}. This design incorporates multiple feeding waveguides, which offer greater design flexibility for the waveguide coupler but also increase the complexity of manufacturing. To achieve a symmetric structure, the device was fabricated by assembling two milled halves of the structure \cite{PhysRevAccelBeams.21.061001}. Two transmission waveguides were placed on opposite sides of the cell to alternately feed the accelerating cavities, with the coupler waveguide connected to the broad wall of the transmission waveguide (E-plane). 

\begin{figure}[!h]
\centering
\includegraphics[width=.8\columnwidth]{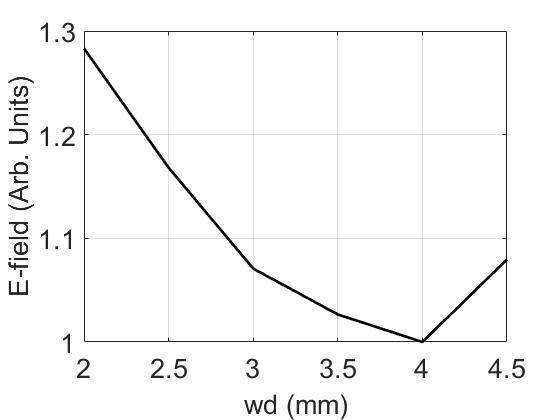}
\caption{The relative maximum surface E-field at different corrugation intervals \textit{$wd$}.}
\label{fig:wd.vs.E}
\end{figure}

\begin{figure*}
\centering
\includegraphics[width=2\columnwidth]{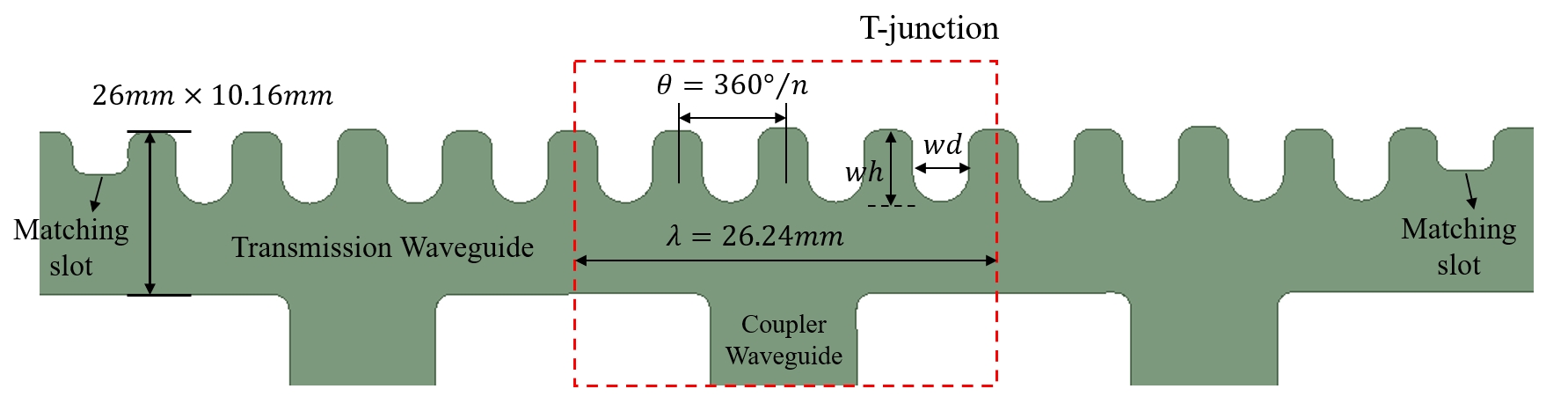}
\caption{The HFSS design of waveguide power feeder: the red dashed box includes a single slow-wave transmission unit. }
\label{fig:hfss_design}
\end{figure*}

The RF phase velocity is primarily determined by the cross-sectional dimensions of the main waveguide and the corrugation parameters of the slow-wave structure (Fig.\ref{fig:hfss_design}). It can be observed that adding more slots in the T-junction unit improves field uniformity in design. However, this improvement is limited by the precision of the machining tools. Based on these considerations, we selected a slot number of \textit{$n = 4$}, and adjusted the slot depth \textit{$wh$} at a specific \textit{$wd$} so that the phase advance per corrugation period approximates \textit{$\pi / 2$}. To balance the impact on the maximum surface E-field (as shown in Fig.~\ref{fig:wd.vs.E}) and the feasible machining range, \textit{$wd$} was set to 3.5 mm. Matching sections were added at both ends of the transmission waveguide to ensure impedance matching with the smooth waveguide. The scattering matrix of the T-junction can be written as

\begin{equation}
S_{3}=\left[\begin{array}{ccc}
0.18 & 0.82 & 0.54 j \\
0.82 & 0.18 & -0.54 j \\
0.54 j & -0.54 j & -0.64
\end{array}\right].
\label{eq:matrix}
\end{equation}

Compared to the smooth waveguide, the wrinkled structure causes a significant change in the impedance for microwaves. The use of matching slots ensures that microwave power is fed into the network with minimal reflection.

\subsection{accelerating cavity design}
\begin{figure}[!h]
\centering
\includegraphics[width=\columnwidth]{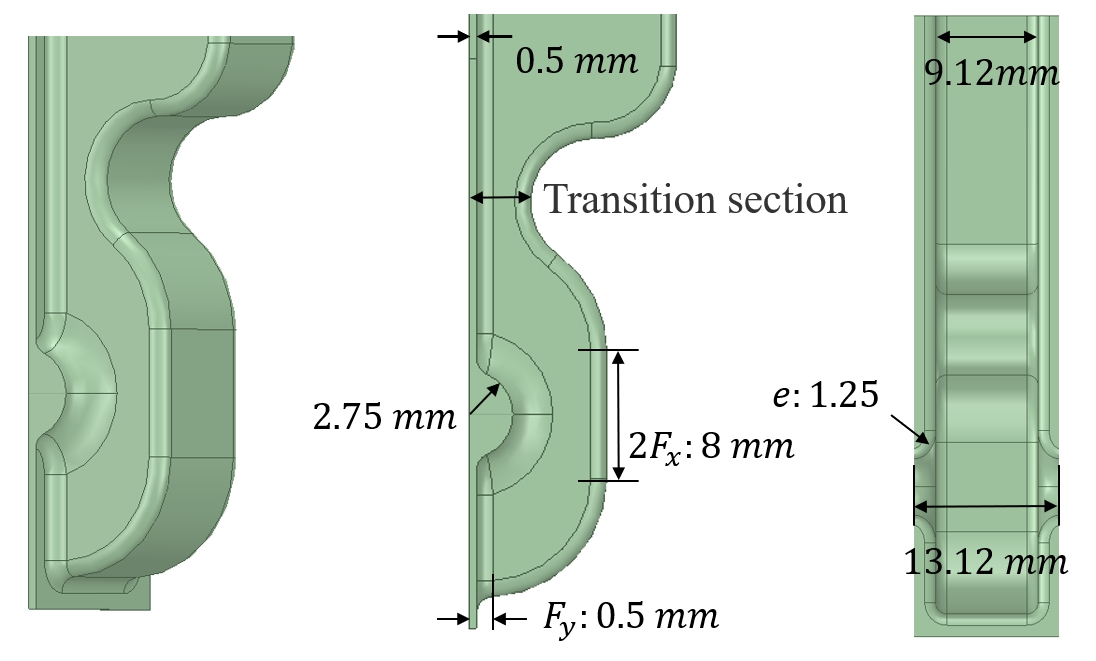}
\caption{The HFSS design of the accelerating cavity with coupler waveguide: elliptical fillets are used at the connection between beam iris and cavity to reduce the surface electric field. The length of the straight line sections in the cell geometries, \textit{$F_x$} and \textit{$F_y$}, are 4 mm and 1.2 mm, respectively.}
\label{fig:cell_shape}
\end{figure}

The geometric design of the accelerating cavity is based on the racetrack-shaped cells of the CLIC-G-OPEN, which are well-optimized and suitable for halves fabrication. Due to limitations in machining accuracy, the cavity shape with nose cones and high shunt impedance was not selected. Fig.~\ref{fig:cell_shape} shows the HFSS model of the identical cell with a coupler waveguide. The accelerating cell is connected to the rectangular waveguide through an arc section. The main parameters are presented in Table.~\ref{tab:para1}.

\begin{figure}[!h]
\centering
\includegraphics[width=\columnwidth]{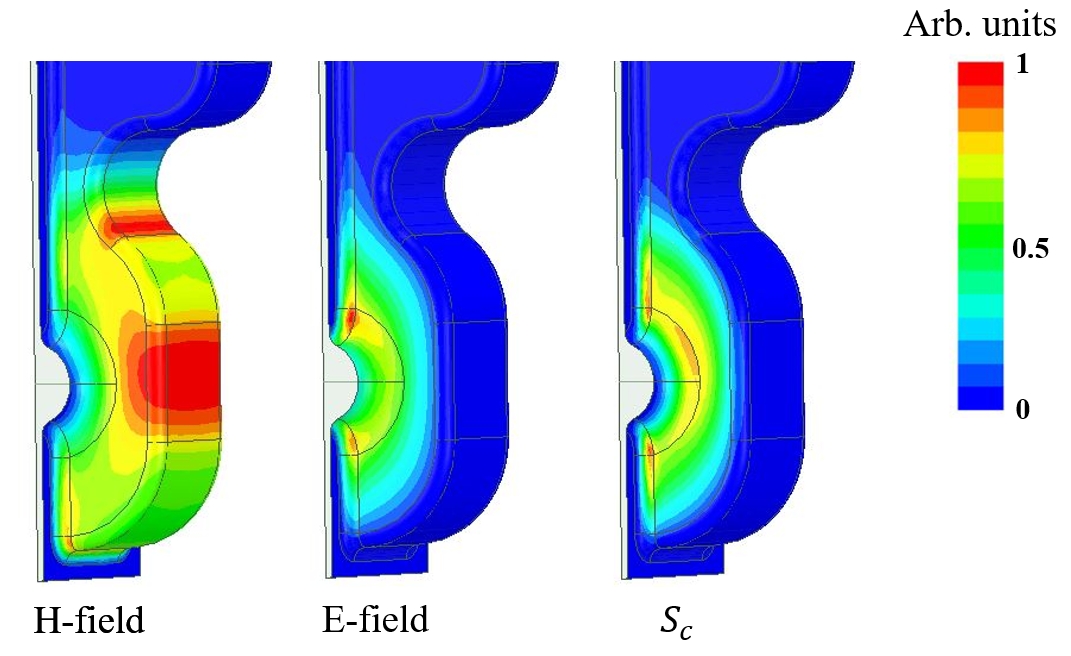}
\caption{Surface field plot of the accelerating cell with transition waveguide.}
\label{fig:field_plot}
\end{figure}

\begin{table}[!htb]
\caption{\label{tab:para1}RF parameters of parallel-coupled single cell.Field values are for an average accelerating gradient of 200 MV/m.}
\begin{ruledtabular}
\begin{tabular}{lc}
\textrm{Parameter}&
\textrm{Parallel-coupled cell}\\
\colrule
Frequency & 11.424 GHz  \\
iris radium & 2.75 mm  \\
\textit{$Q_0$} & 8200  \\
Shunt impedance & 89 $\mathrm{M\Omega /m}$ \\
Peak surface E & 257 MV/m \\
Peak surface H with transition section & 0.57 MA/m \\
Maximum \textit{$S_{c}$} & 32 $\mathrm{MW / mm^2}$ \\
\end{tabular}
\end{ruledtabular}
\end{table}

The coupling coefficient is adjusted by the size of the transition section opening, where the peak magnetic field appears, as shown in Fig.~\ref{fig:field_plot}. The peak electric field and modified Poynting vector (\textit{$S_c$}) are concentrated near the beam hole, consistent with axis-coupled calculation results. Furthermore, based on previous test results showing no noticeable increase in BDR for side-coupled cells compared to axis-coupled ones \cite{dolgashev2011status}, this cell type is suitable for high-gradient parallel-coupled structures. During the design process, we observed that the amplitude of the electric field in each cavity is unequal when the resonant frequency of all cells is kept the same. The size of the beam aperture in this design is comparable to that of the CLIC-G-OPEN TW structure, making the inter-cavity coupling non-negligible. To address this issue, simulations were performed to optimize the resonant frequency difference between the side cell and the middle cell, which was set to 9 MHz to ensure an equal-amplitude E-field distribution at the single resonance frequency of the structure.

\subsection{Design and analysis of the structure}
Similar to the optimization of filling time for travelling-wave accelerating structure\cite{2008RF}, we calculated the optimal filling time for parallel-coupled structure, expressed as:

\begin{equation}
t_{f} \simeq \frac{2.512 Q_{0}}{\omega\left(1+\beta^{\prime}\right)}
\label{eq:filltime}
\end{equation}

\begin{figure}[!h]
\centering
\includegraphics[width=0.9\columnwidth]{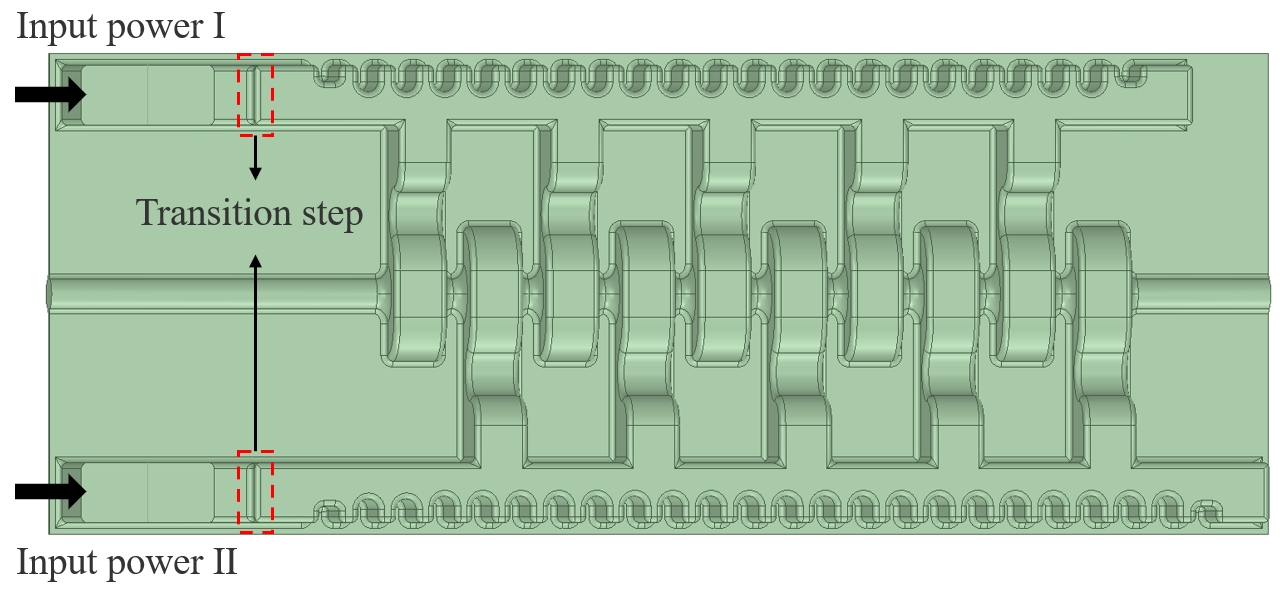}
\caption{Half geometric design of slow-wave parallel-coupled accelerating structure with 10 cells.Transition step is used to connect waveguides in different broadwall size.}
\label{fig:hlave_model}
\end{figure}

At an input pulse width of 40 ns,  the optimized coupling coefficient of the structure (\textit{$\beta^{\prime}$}) and a single cell (\textit{$\beta$}) were determined to be approximately \textit{$\beta^{\prime}\approx6.2$} and \textit{$\beta\approx4$}\cite{jiang}. Two input RF microwaves with equal power and a $180^{\circ}$ phase difference were fed into the full structure, shown in Fig.~\ref{fig:hlave_model}. A regular waveguide with a relatively short broadwall length was adopted to inject power into the corrugated waveguide network, avoiding the growth of undesired microwave modes and minimizing efficiency loss.

\begin{figure}[!h]
\centering
\includegraphics[width=0.9\columnwidth]{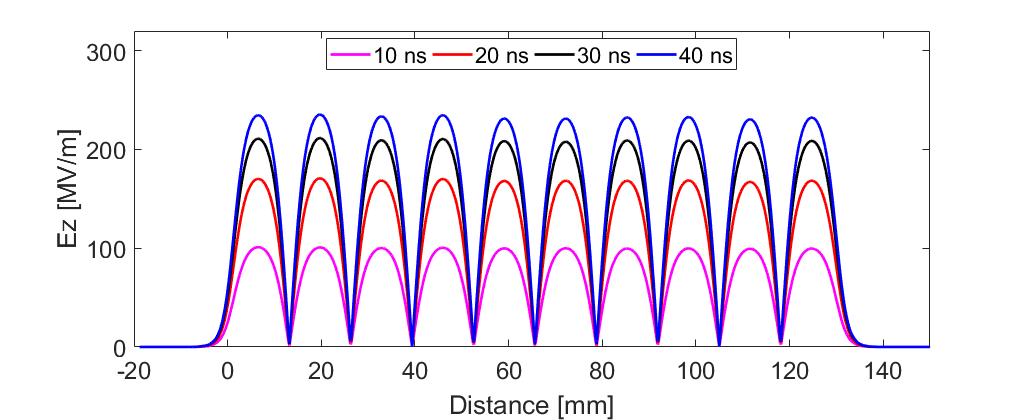}
\caption{The electric field distribution along the axis at different times, with a pulse of 100 MW.}
\label{fig:real_Ez_time}
\end{figure}

\begin{figure}[!h]
\centering
\includegraphics[width=0.9\columnwidth]{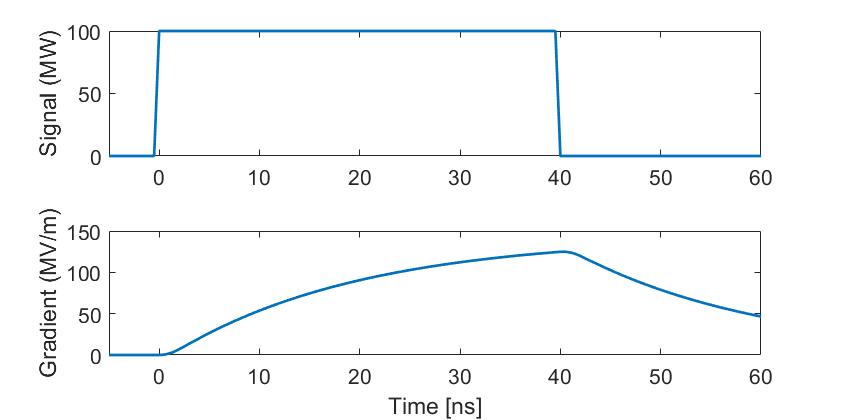}
\caption{The temporal average gradient of structure with a pulse of 100 MW and 40 ns of length. The rising time was set to be 1 ns}.
\label{fig:gradient_time}
\end{figure}

For short-pulse accelerating structures, time-domain analysis tools are essential to simulate the real-time distribution of the electromagnetic field. A method based on the fast Fourier Transform was developed and validated to provide accurate real-time field data \cite{jiang}. Using this method, we analyzed the filling process of the designed parallel-coupled structure. With an input pulse of 100 MW, a 40 ns pulse length, and a 1 ns rising time, the simulation results showed that power begins filling each cell simultaneously at $t = 4 \mathrm{ns}$, and the E-field in each cell remains consistent over time. Fig.~\ref{fig:real_Ez_time} and Fig.~\ref{fig:gradient_time} demonstrated the \textit{$E_{z}$} field value distribution along the axis and the temporal average accelerating gradient, respectively. The parallel-coupled structure was designed for single-bunch input, making it compatible with pulses of different shapes.

\section{\label{tuning}Fabrication, tuning, and low power test analysis}

Based on the RF design of the parallel-coupled structure, we provide a detailed description of the fabrication process in the first subsection. The second subsection presents the cold test and tuning results of the structure. Due to the power coupling between adjacent cavities through the beam holes, amplitude distribution measurements of the electric field become challenging. However, for structures operating under short pulses, the accuracy of the single-cavity frequency is more meaningful than steady-state electric field distribution when considering the effects of high-power testing. In the third subsection, we present a coupling circuit model for the parallel-coupled structure, explaining that the bead-pull results do not correspond to the true on-axis electric field distribution within the structure.

\subsection{Fabrication}

\begin{figure}[!htb]
\centering
\includegraphics[width=0.9\columnwidth]{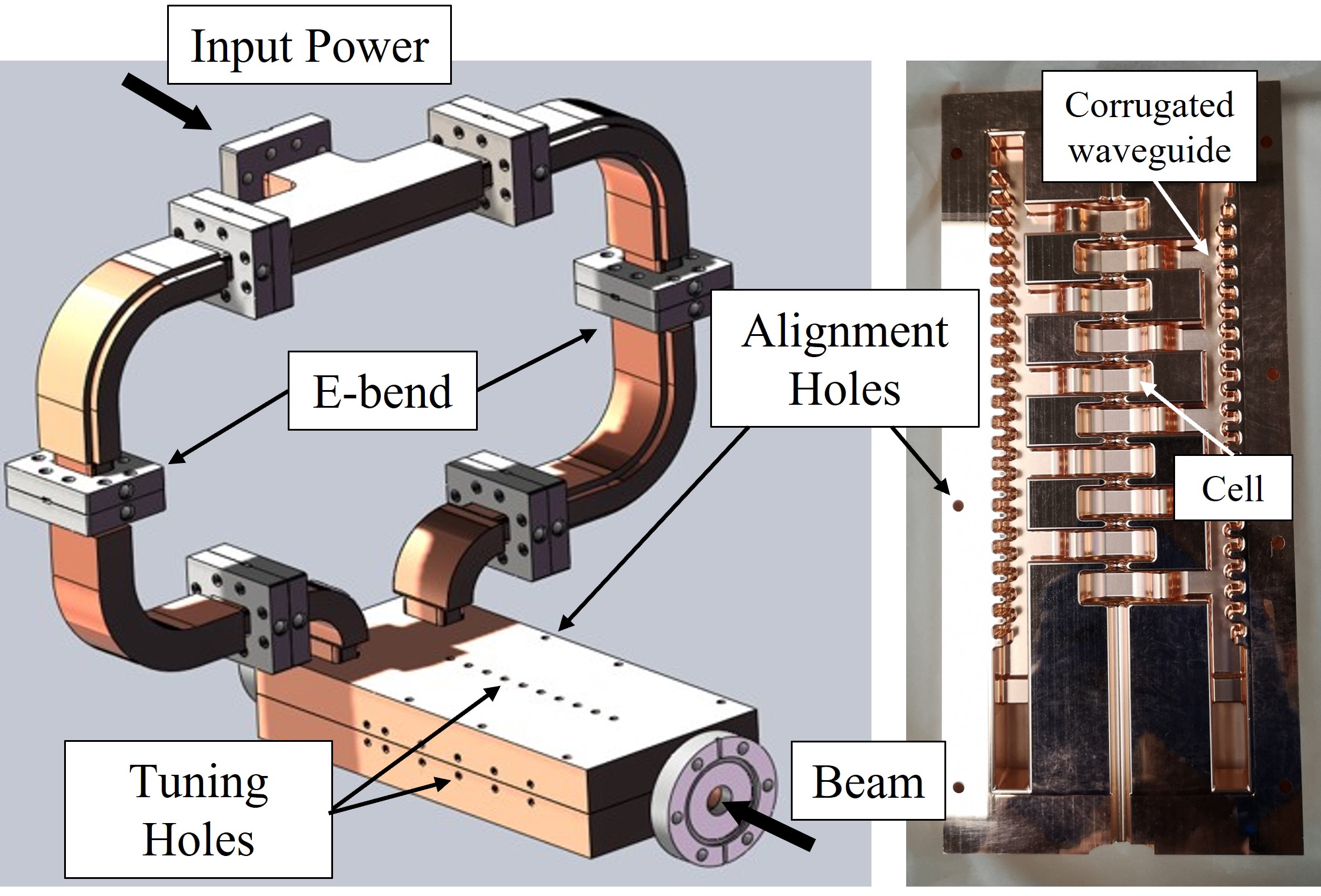}
\caption{Illustration of the mechanical design of the parallel-coupled structure with E-bend waveguide and power splitter.}.
\label{fig:mech_design}
\end{figure}

The mechanical design of the slow-wave parallel-coupled structure consists of upper and lower copper blocks, as shown in Fig.~\ref{fig:mech_design}. Each part is machined according to the half structure split by the middle cross-section along the long dimension of the slow-wave waveguide. The use of two halves spliced together effectively reduces the complexity of fabrication but imposes high requirements on positioning accuracy. Therefore, four alignment holes are provided on each side of the copper block for alignment during welding, and tuning holes are located on the side walls and top.

To simplify the fabrication process, no water-cooling channels are incorporated within the structure. Temperature regulation and stabilization of the structure would be achieved through externally attached water-cooling strips in high-power testing. The maximum error for the transmission waveguide dimensions was -0.01 mm, corresponding to a phase shift of $4^{\circ}$. On the other hand, the error for the accelerating cavity was -0.015 mm, corresponding to a frequency deviation of approximately 15 MHz.

\subsection{Tuning and low-power test}

The structure was tested under low-power input conditions, and the measurements include cavity frequency, overall S-parameters of the structure, and field distribution. To minimize errors between two ports of the vector network analyzer, the final structure was connected to a power splitter for single-port testing, shown in Fig.~\ref{fig:mech_design}.

\begin{figure}[!htb]
    \centering
    \subfigure[\label{fig:S_parameter_a}]{\includegraphics[width = 0.45\columnwidth]{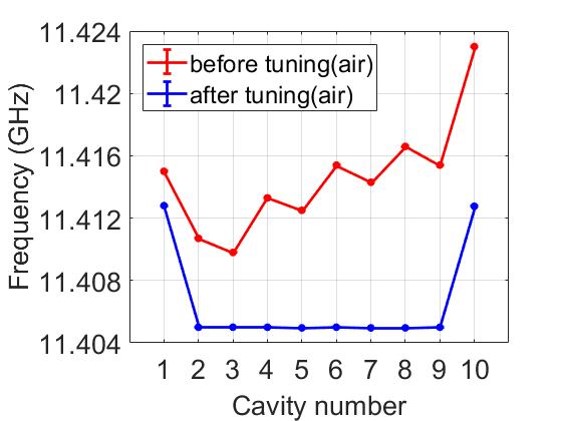}}
    \subfigure[\label{fig:S_parameter_b}]{\includegraphics[width = 0.45\columnwidth]{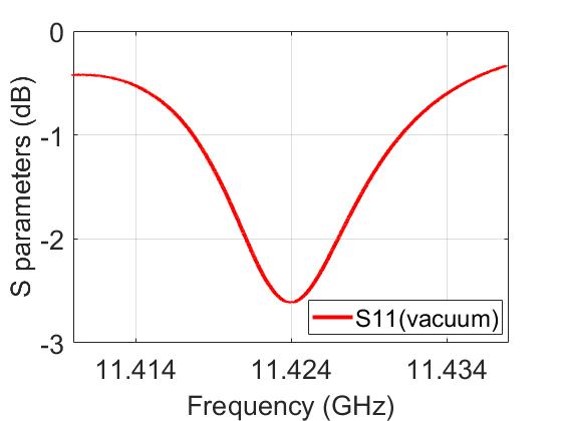}}\\
    \subfigure[\label{fig:S_parameter_c}]{\includegraphics[width = 0.45\columnwidth]{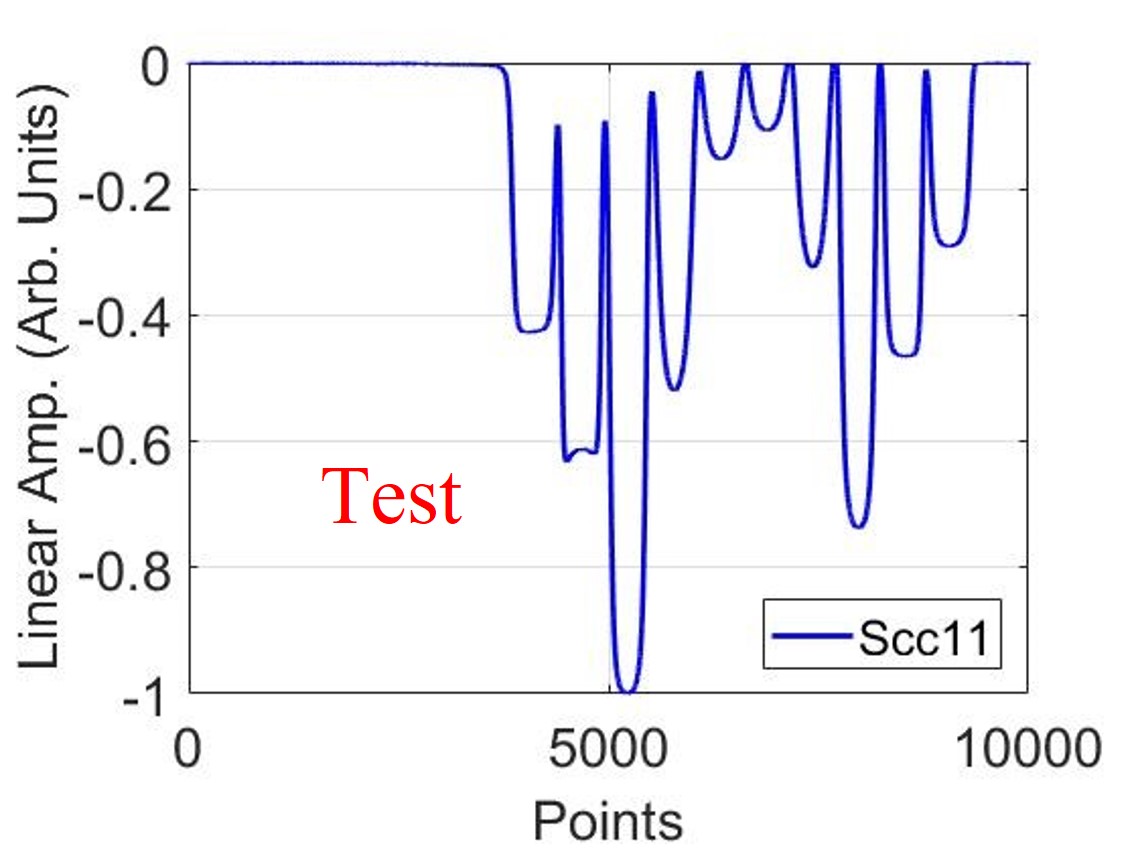}}
    \subfigure[\label{fig:S_parameter_d}]{\includegraphics[width = 0.45\columnwidth]{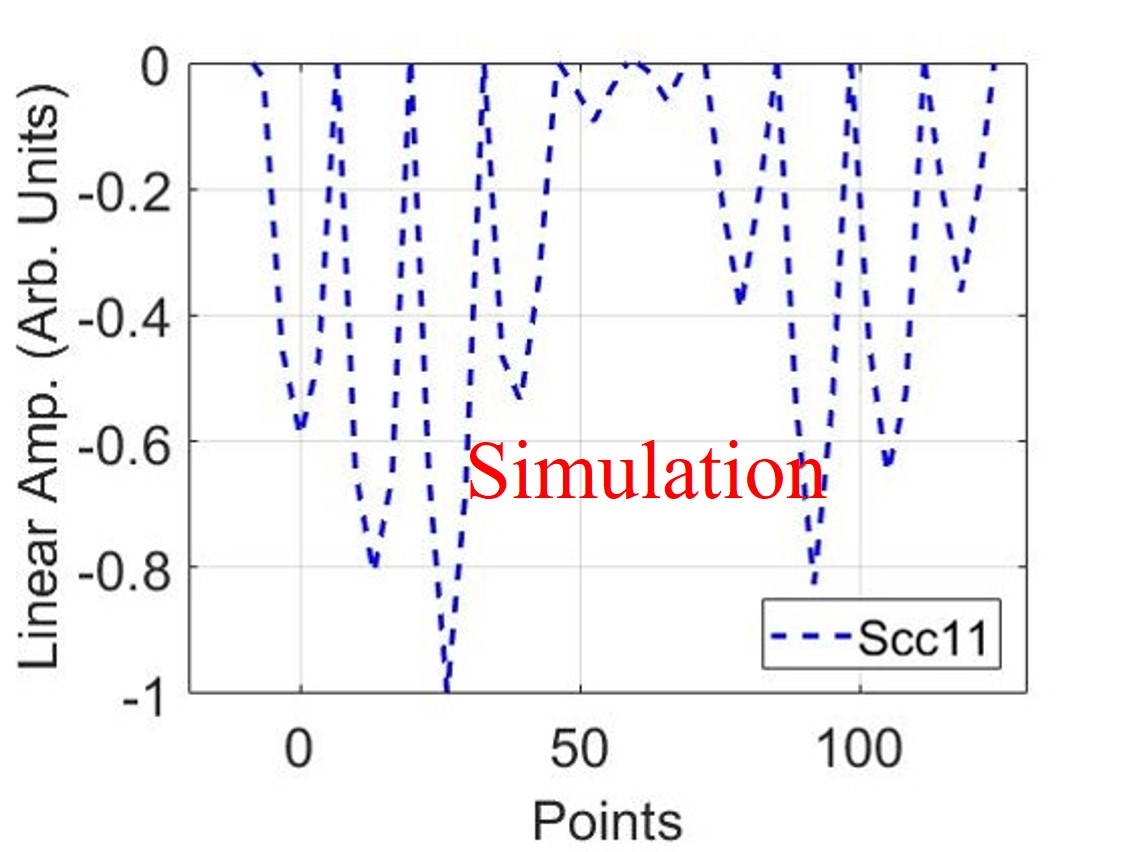}}
    \caption{Tuning results of the slow-wave parallel-coupled structure. (a) and (b) show the resonant frequencies of the tuned cells and the structure, respectively. (c) and (d) are the test and simulation bead-pull data of the structure.}
    \label{fig:S_parameter}
\end{figure}

Unlike conventional multi-cavity accelerating structures, the fields in the cells of the parallel-coupled structure are excited by coupling with distributed waveguides rather than through cell-to-cell coupling. Consequently, the S-parameter curve of the structure is expected to exhibit a single peak. Fig.~\ref{fig:S_parameter_a} shows the single cavity frequencies before and after tuning. Before tuning, the single-cavity frequency of the structure exhibits a distribution from low to high, due to the counterweight being placed closer to the tail during welding, resulting in tighter contact surfaces. Tuning was accomplished by shorting all cells except the one under test with two conductive bars inserted from opposite sides of the iris. The overall frequency of the structure was tuned to 11.424 GHz, with a frequency difference of less than 40 kHz between the middle cells. For the first and last cavities, different resonance frequencies were necessary to ensure the S-parameter curve exhibited a single peak. It should be noted that the discrepancy between the measured results and the design value is attributed to machining errors in the beam apertures, which also highlights that the cavities cannot be considered completely isolated from one another. The S-parameter of the structure (Fig.~\ref{fig:S_parameter_b}) was measured using a power splitter, which indicated that the coupling coefficient of the structure is approximately 6.7, meeting the design specifications for operation under short pulses.

We used the non-resonant perturbation bead-pull method \cite{1966A} to measure the on-axis field. In this case, a metal bead induces a disturbance to the electric field at its location, which leads to a change in the reflection coefficient of the structure. Therefore, by moving the bead at a constant speed along the structure, we obtained an unexpected curve of the relative change in the reflection coefficient versus the bead position, as shown in Fig.~\ref{fig:S_parameter_c}. In HFSS, a copper sphere with a diameter of 1 mm is placed and simulated at different positions along the structure, which is illustrated in Fig.~\ref{fig:S_parameter_d}. Theoretically, the cavities working in \textit{$\pi$} mode are considered to be isolated from each other, resulting in an equivalent change in the reflection coefficient when the single cell is detuned. However, a non-flat amplitude distribution was observed in both tests and simulations, indicating that the bead-pull method is unsuitable for field measurement in this case. We attribute these results to inter-cavity coupling and propose a circuit model of the parallel-coupled structure for further analysis.

\subsection{Coupling circuit model}
According to the circuit model of a resonant cavity, the energy stored in the cavity \textit{$U_\text{in}$} is proportional to the external power input \textit{$P_\text{ex}$}, with the ratio mainly determined by the reflection coefficient \textit{$\Gamma$}:
\begin{equation}
    \frac{Q_{0}}{\omega_{0}} U_{\text{in}} = P_{\text{in}}\left(1-\Gamma^{2}\right)
    = P_{\text{in}} \frac{4 \beta}{(1+\beta)^{2} + \left( \frac{2 Q_{0} \Delta \omega}{\omega_{0}} \right)^{2}}
    \label{eq:cell_circuit}
\end{equation}
where \textit{$\omega_{0}$}, \textit{$\Delta \omega$}, \textit{$Q_{0}$} and \textit{$\beta$} are the angular frequency of input RF power, frequency difference between the input signal and the resonant frequency, unloaded quality factor, and coupling coefficient, respectively.

The voltage relationship between the cell and the external source is given by the root of Eq.~\ref{eq:cell_circuit}, where voltage can be represented as a vector on the complex coordinate system. The accelerating voltage \textit{$V_\text{n}$} could be described as a function of the external source voltage \textit{$I_\text{n}$}:
\begin{equation}
    \sqrt{\frac{\omega_{0}}{Q_{e}}} I_{\text{n}} = V_{\text{n}} \sqrt{\frac{\omega_{0}^{2}}{4 Q_{L}^{2}} + (\Delta \omega)^{2}}
    = V_{\text{n}} \left( \pm \Delta \omega \pm \frac{\omega_{0}}{2 Q_{L}} j \right)
    \label{eq:cell_voltage}
\end{equation}
where \textit{$Q_\text{L}$}, \textit{$Q_\text{e}$} are the loaded and external quality factors. For simplicity, we adopted \textit{$\Delta \omega$} and \textit{$\frac{\omega_{0}}{2 Q_{L}}$} as the absolute values of real and imaginary parts of the coefficient vector, respectively. The electromagnetic power dissipated in the external load corresponds to the electrical signal defined as \textit{$\frac{j\omega_{0}}{2 Q_{L}} V_{n}$}. The reflected signal, due to the impedance mismatch between the input waveguide and the cavity resonant circuit, can be expressed as:
\begin{equation}
    R_{n} = \sqrt{\frac{\omega_{0}}{Q_{e}}} I_{\text{n}} - \frac{j \omega_{0}}{2 Q_{L}} V_{n}
    \label{eq:reflect_signal}
\end{equation}

Based on these definitions, we deduced that the real and imaginary parts in Eq.~\ref{eq:cell_voltage} should be negative and positive, respectively. Considering the coupling behavior between adjacent cells in the parallel-coupled structure, an equation is established that reflects the coupling mechanism between the multi-cells and waveguides:
\begin{widetext}
\begin{equation}
    \left[ \begin{array}{cccc}
    - \Delta \omega_{1} + j \frac{\omega_{0}}{2 Q_{L}} & k & \cdots & 0 \\
    k & - \Delta \omega_{2} + j \frac{\omega_{0}}{2 Q_{L}} & k & \vdots \\
    \vdots & \cdots & \ddots & k \\
    0 & \cdots & k & - \Delta \omega_{n} + j \frac{\omega_{0}}{2 Q_{L}}
    \end{array} \right]
    \left[ \begin{array}{c}
    V_{1} \\
    V_{2} \\
    \vdots \\
    V_{n}
    \end{array} \right]
    = \sqrt{\frac{\omega_{0}}{Q_{e}}}
    \left[ \begin{array}{c}
    I_{1} \\
    I_{2} \\
    \vdots \\
    I_{n}
    \end{array} \right]
    \label{eq:coupled_matrix}
\end{equation}
\end{widetext}
where \textit{$k$} is the coupling constant. This constant satisfies the condition that one of the eigenvectors of the coupling matrix aligns with the distribution of electric field amplitude in the \textit{$\pi$} mode. Eq.~\ref{eq:coupled_matrix} allows us to calculate the theoretical frequency difference between the end cavity and the middle cavity, ensuring uniform electric field amplitude. For the parallel-feeding power distribution network, the reflected power will flow back into the network and affect other cells while it is not critically coupled. In conjunction with the scattering matrix of the RF feed system \cite{jiang}, the incident signal of all cells could be calculated as:
\begin{equation}
    \boldsymbol{S}_{n+1}
    \left[ \begin{array}{c}
    a_{0} \\
    R_{1} \\
    \vdots \\
    R_{n}
    \end{array} \right]
    = \left[ \begin{array}{c}
    b_{0} \\
    \sqrt{\frac{\omega_{0}}{Q_{e}}} I_{1} \\
    \vdots \\
    \sqrt{\frac{\omega_{0}}{Q_{e}}} I_{n}
    \end{array} \right]
    \label{eq:network}
\end{equation}
where \textit{$a_{0}$} is the amplitude of the input signal, \textit{$b_{0}$} is the reflected one, and \textit{$\boldsymbol{S}_{n+1}$} represents the scattering matrix of the network with N ports for cells and one port for the power source.

\begin{figure}[!h]
    \centering
    \subfigure[The normalized variation of S11 vs position of perturbation\label{fig:coupled_a}]{\includegraphics[width = 0.8\columnwidth]{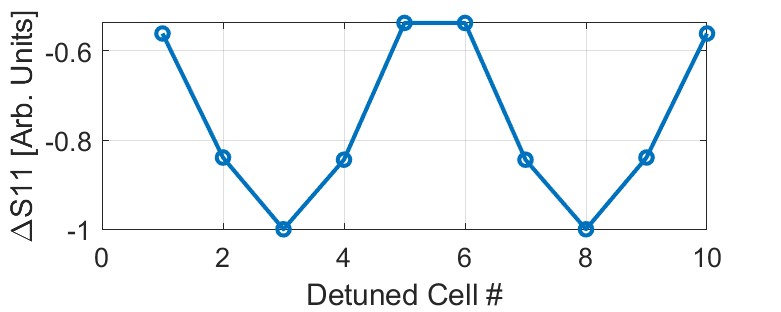}}
    \subfigure[On-axis normalized E-field amplitude\label{fig:coupled_b}]{\includegraphics[width = 0.8\columnwidth]{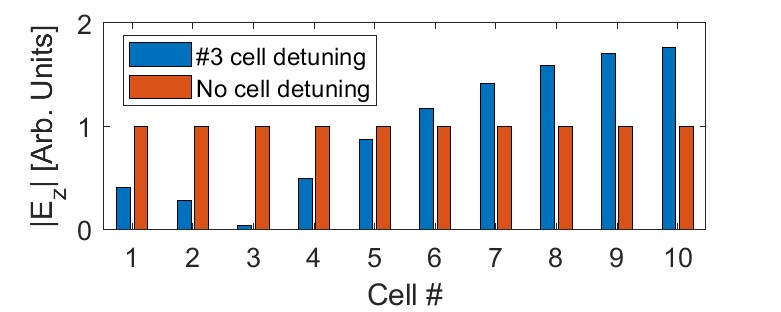}}
    \caption{Theoretical calculation results of the circuit model for the parallel-coupled structure.}
    \label{fig:cell_coupled}
\end{figure}

\begin{figure*}
\centering
\includegraphics[width=2\columnwidth]{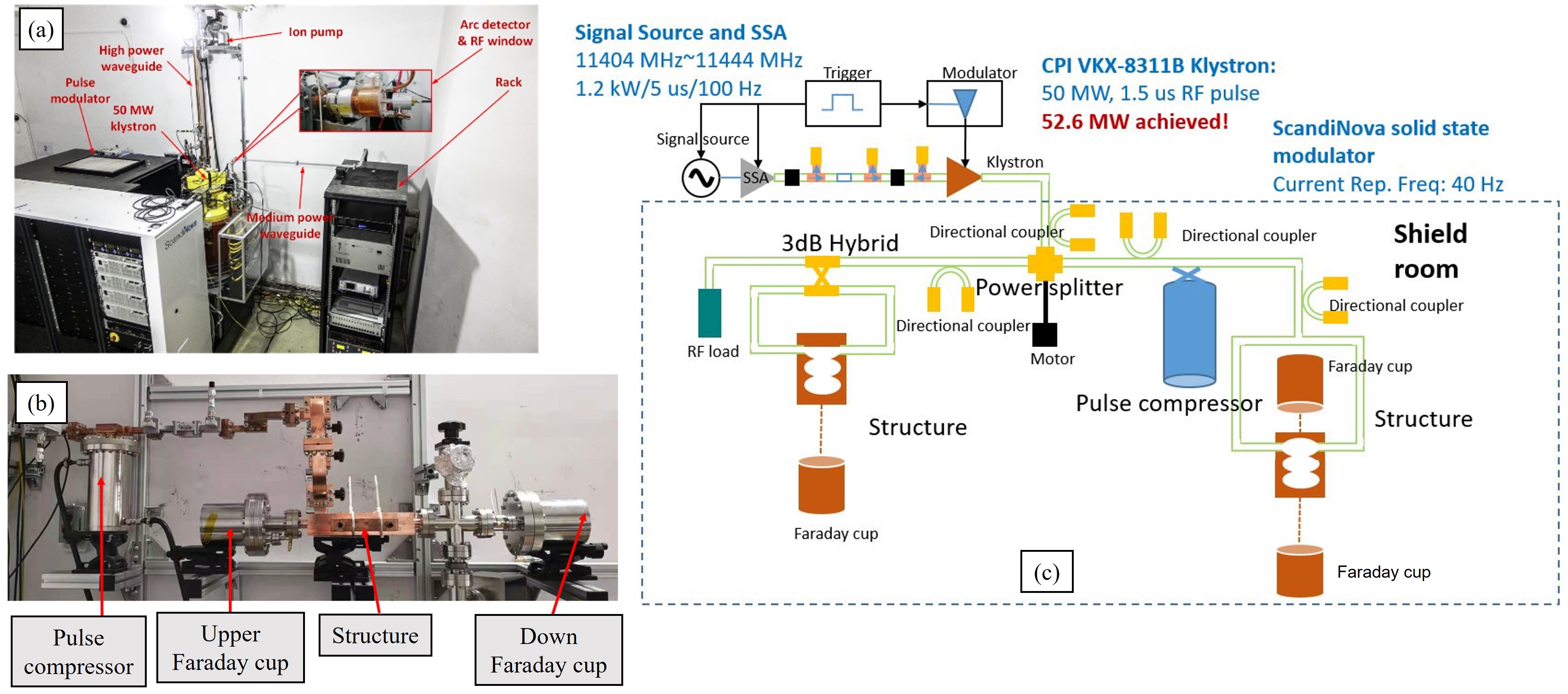}
\caption{Setup of high power test platform. (a) Power supply system outside the shield room. (b) Main test section of structure. (c) The layout figure of TPOT-X. }
\label{fig:TPOT}
\end{figure*}

In the non-resonant perturbation theory, the resonant frequency change resulting from the conducting prolate spheroid placed in cells leads to a change in the reflection coefficient. Choosing the input signal \textit{$a_{0}=1$}, we can derive the variation curve of the reflected coefficient as the detuned cell varies, which shows a high degree of similarity with the experimental results in Fig.~\ref{fig:coupled_a}. The normalized E-field distributions in the parallel-coupled structure, with no cell and one cell detuning, were illustrated in Fig.~\ref{fig:coupled_b}.

\section{\label{test}High power test}
\begin{figure}[!h]
\centering
\includegraphics[width=\columnwidth]{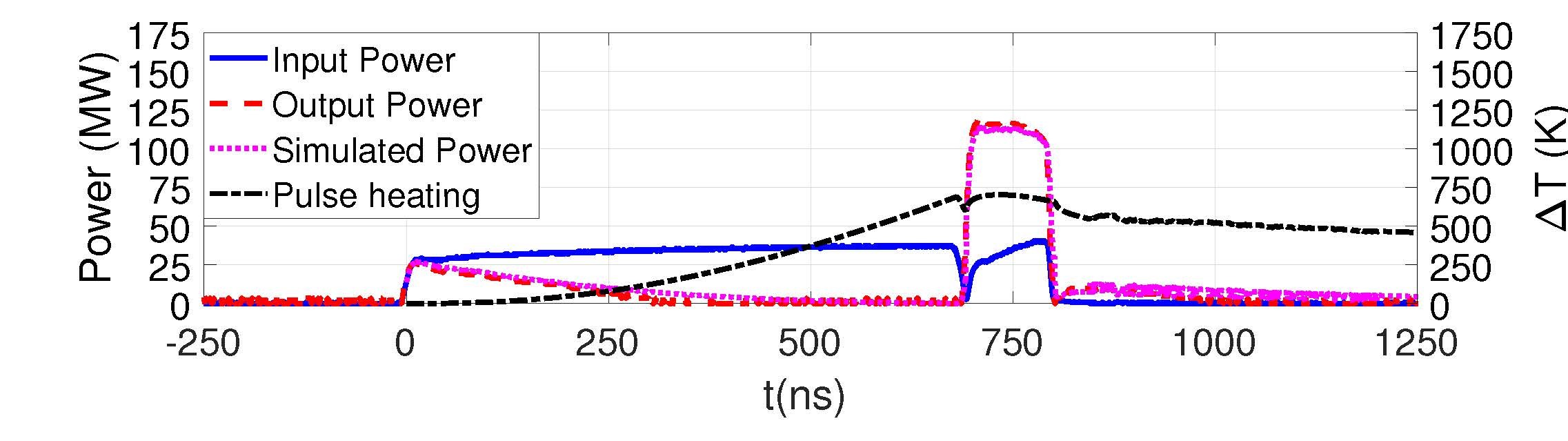}
\caption{After amplitude modulation, the output of pulse compressor is approximated as a square pulse of 40ns.}.
\label{fig:in_waveform}
\end{figure}

The slow-wave parallel-coupled accelerating structure was tested at the Tsinghua high power test stand for X-band (TPOT-X). Fig.~\ref{fig:TPOT} depicts the layout of the high-power test platform, with the main test section inside the shield room and the power supply system outside. We used a pulse compressor after the klystron to generate ultrashort pulses that exceed the klystron limit. Smooth or corrugated spherical or cylindrical resonant cavities with a quality factor \textit{$Q_{0} > 100000$} are commonly used as pulse compressors\cite{compressor1,compressor2,compressor3}. Consequently, we designed and fabricated an X-band corrugated cylinder pulse compressor, with the quality factor \textit{$Q_{0} \simeq 117000$} and coupling coefficient \textit{$\beta \simeq 3.67$}. Fig.~\ref{fig:in_waveform} shows the compressed square pulse waveform with an adjustable width of 40 ns and a power level of 120 MW which could be stably generated by the compressor.

\begin{figure}[!h]
\centering
\includegraphics[width=1\columnwidth]{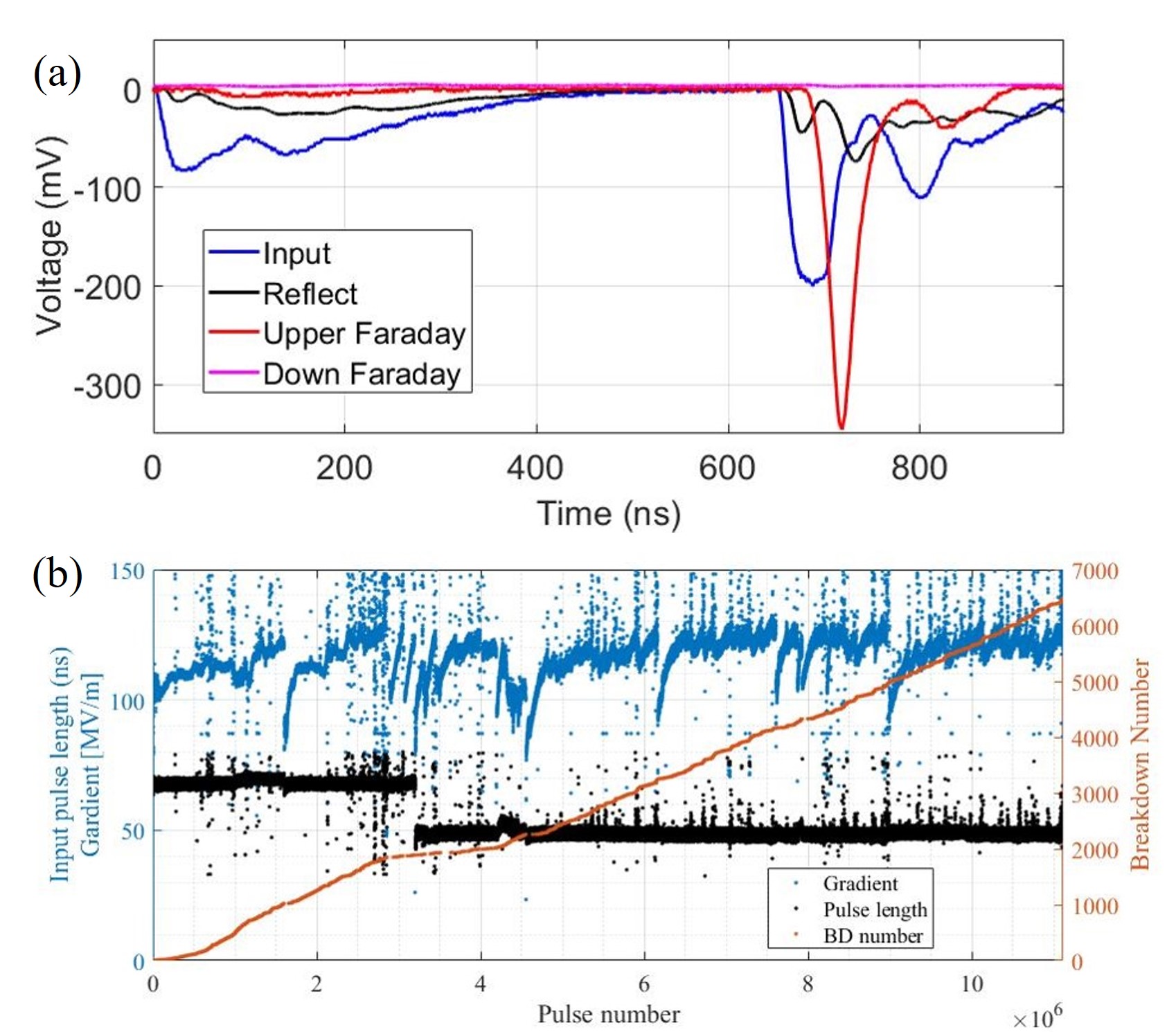}
\caption{(a) For the over-coupled structure, the reflect signal arises as the input power is fed. The dark current signals in upstream and downstream Faraday cups are also monitored. (b) The high power test history of parallel-coupled structure.}.
\label{fig:waveform}
\end{figure}

During the high power RF conditioning, the input and output power waveforms of the structure are recorded to calculate the accelerating gradient, which has been proved to be accurate. Power waveforms at waveguide network joints were acquired and diagnosed by installing directional couplers and crystal detectors, allowing measurement of the incident and reflected RF signals. A typical waveform during the test is shown in Fig.~\ref{fig:waveform}. Dark current signals from the upstream Faraday cup were used as the criterion for RF breakdown events, which typically occur at the end of input pulses. A threshold detector on RF signals in the real-time measurement system was used to trigger and collect the breakdown events. The power level increased step-by-step with a 30-second growth period, during which the power level remained constant. If no breakdown occurs in the growth period, the power level increased by 0.5 MW. On the contrary, the microwave input was turned off for 30 seconds, and the power level decreased by 0.2 MW at the next start of the period. 

\begin{table*}
    \caption{\label{tab:par_comp}High-power test results of four X-band accelerating structures.}
    \begin{ruledtabular}
    \begin{tabular}{lcccc}
         \textrm{Structure}&
\textrm{Pulse width}&
\textrm{Gradient\footnote{The accelerating gradient within \textit{$1\times10^{7} $} conditioning pulses }}&
\textrm{Maximum gradient\footnote{The final achieved accelerating gradient. The total high-power pulse number for CLIC-G is \textit{$1\times10^{8} $} and the total pulse number for structures from top to bottom in the table are \textit{$1.1/2.5/3.2\times10^{7} $}, respectively}}&
\textrm{BDR\footnote{The breakdown rate at the maximum gradient}}\\
\colrule
        \textrm{Parallel-coupled slow-wave structure} & 47 ns & 120 MV/m & 130 MV/m & {$5\times10^{-4} $}\\
        \textrm{Tsinghua TW single-cell structure} & 50 ns & 110 MV/m & 123 MV/m & {$3\times10^{-4} $}\\
        \textrm{Tsinghua TW 10-cell structure} & 90 ns & 80 MV/m & 93 MV/m & {$2\times10^{-5} $}\\
        \textrm{CLIC-G-OPEN TW 24-cell structure} & 200 ns & 50 MV/m & 100 MV/m & {$6\times10^{-6} $}\\
    \end{tabular}
    \end{ruledtabular}
\end{table*}

The final high-power test history of the parallel-coupled structure is illustrated in Fig.~\ref{fig:waveform}, which displays the total breakdown number and counted pulse. In the early stages of the experiment, the pulse length was set to 70 ns to speed up the processing, with a total of \textit{$3\times10^{6} $} pulses, achieving an accelerating gradient of 120 MV/m. The width of the subsequent \textit{$8\times10^{6} $} pulses was adjusted to the design value, with the tested gradient reaching 130 MV/m for short input power pulses. The estimated breakdown rate was \textit{$5\times10^{-4} $} and \textit{$1\times10^{-3} $}, for the pulse width of 47 ns and 70 ns, respectively.

By the end of high-power testing the parallel-coupled slow-wave structure had undergone \textit{$1.1\times10^{7} $} pulses, which is fewer than most high-gradient accelerating structures. As illustrated in the experimental microwave layout in Fig.\ref{fig:TPOT}, the reflected power flows through the pulse compressor to the klystron power system, rather than being directed to the RF load. Once the structure's gradient reached 100 MV/m, reflected power exceeding 10 MW could potentially cause cumulative damage to the power source system as conditioning continued. Therefore, considering the safety and longevity of the power source, we did not conduct extended high-power testing on the parallel-coupled structure. In future plans, a 3 dB directional coupler will be used to combine and transmit the reflected signals from two identical structures to a steel load, while a two-stage compressor will generate pulses with over 200 MW power.

Table.~\ref{tab:par_comp} shows the high-power test results of four different X-band accelerating structures. The fabrication methods of the Tsinghua 10-cell structure and CLIC-G-OPEN are similar to that of the parallel-coupled slow-wave structure, but the width of the high-power input pulses differs. Compared to the single-cell structure, the parallel-coupled design achieves a comparable accelerating gradient within the same high-power test duration and therefore can be considered equivalent to conditioning 10 single cavities simultaneously. Additionally, among the three multi-cell structures, the potential to reach higher gradients is validated under short-pulse input operation.

\begin{figure}[!h]
    \centering
    \subfigure[Endoscopic images of the right side of cells at different positions of the structure.\label{fig:BD_a}]{\includegraphics[width = 0.9\columnwidth]{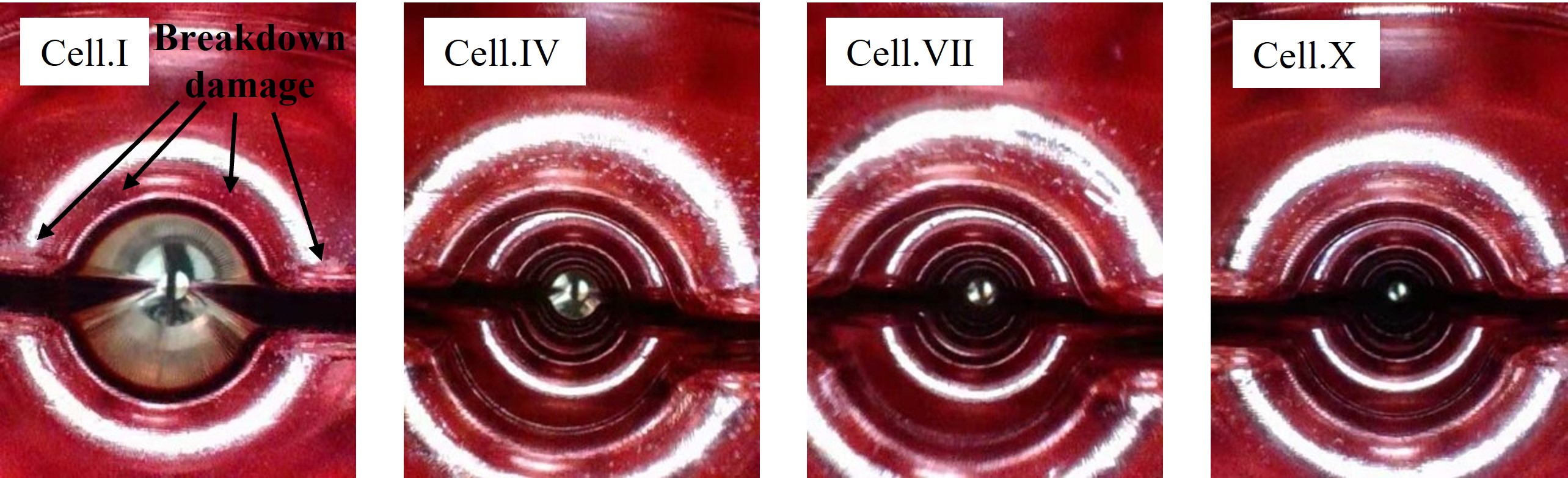}}\\
    \subfigure[Breakdown induced spotty damage distribution along the structure\label{fig:BD_b}]{\includegraphics[width = 0.9\columnwidth]{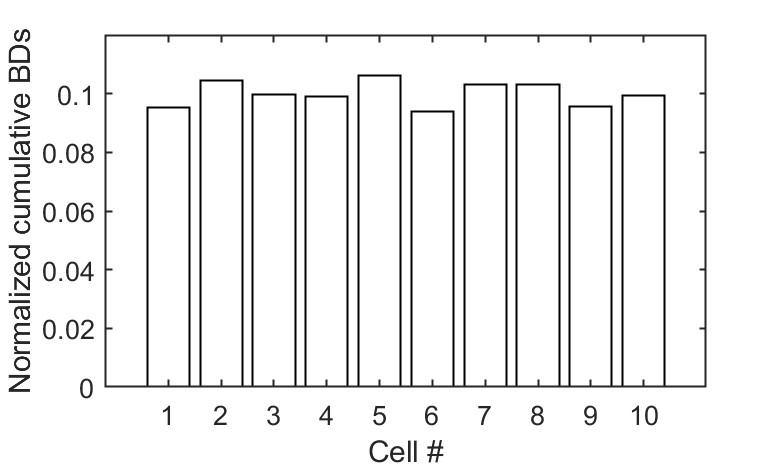}}
    \caption{Endoscopic images and breakdown distribution inside the structure.}
    \label{fig:inner_pic}
\end{figure}

The S-parameters were remeasured to evaluate the effects in RF properties due to the high-power testing,  with no significant changes observed. Subsequently, an endoscope was used to inspect the breakdown sites inside the tested structure, with white spots shown in Fig.~\ref{fig:inner_pic} marking the breakdown locations. In contrast to most other high gradient structures, the normalized cumulative breakdown number shown in Fig.~\ref{fig:inner_pic} demonstrates a uniform breakdown distribution without peak value along the parallel-coupled structure, which confirms the similar distribution of electric field strength. The spot densities of the two opposing iris areas were found to be similar after comparing different side images. Breakdowns primarily occurred near the beam hole and central gap plane, consistent with the distribution of the maximum surface modified Poynting vector (\textit{$S_c$}).

\section{\label{conclusion}Conclusion}

In this paper,we designed and fabricated the first X-band short-pulse slow-wave parallel-coupled accelerating structure and conducted tests on it. To ensure phase synchronization, we selected a corrugated waveguide as the power feeding network to reduce the phase velocity. The structure, which delivers power simultaneously and independently to each cavity, helps to effectively reduce the field-building time, thus enabling short-pulse operation. Simulation results showed that the structure can operate with ultra-short pulses of 40 ns, with electric field amplitude variations in each cell being equal and changing synchronously. This demonstrates its potential for high-gradient operation. Due to the structural complexity, we used a manufacturing technique where two copper blocks were used to machine half of the structure, which were then joined together. Moving on, we described the cold testing and tuning results of the structure, where we found that beam-hole coupling still existed and affected the overall reflection coefficient. Non-resonant perturbation methods could not be used to measure the actual electric field amplitude distribution in the structure. The multi-cell coupling circuit for the parallel-feeding structure is thus proposed, and the accuracy is verified by calculations. Finally, we conducted the high-power test on the TPOT-X platform, which is based on a corrugated cylinder pulse compressor, and achieved an accelerating gradient of 120 MV/m with a working pulse width of 47 ns at the BDR level of \textit{$5\times10^{-4} $}. After conditioning, endoscopic images revealed no significant damage to the structure, and the field strength across the cavities was uniform. In conclusion, the parallel-coupled structure is a promising choice for multi-cavity short-pulse accelerators and provides an effective solution for future high-gradient short-pulse applications.

\nocite{*}

\bibliography{a_couple}

\end{document}